\documentclass[amssymb,amsmath,twocolumn, prb]{revtex4}
\usepackage{graphicx}
\usepackage{subfigure}
\usepackage{tikz}
\usepackage{bm}
\begin{document}
\setcounter{page}{1}
\title[]{
Phase transitions and spectral properties of the ionic Hubbard model in one dimension
}
\author{Ara \surname{Go}}
\affiliation{Department of Physics and Astronomy, Seoul National University, Seoul 151-747, Korea}
\author{Gun Sang \surname{Jeon}}
\email{gsjeon@ewha.ac.kr}
\thanks{FAX: +82-2-3277-2372}
\affiliation{Department of Physics, Ewha Womans University, Seoul 120-750, Korea}
\date{\today}

\begin{abstract}
The ionic Hubbard model is investigated at half filling at zero temperature.
We apply the cellular dynamical mean-field theory to the one-dimensional 
ionic Hubbard model to
compute local quantities such as double occupancy and staggered
charge density. Both quantities provide general transition behavior 
of the model from a band insulating phase to a Mott insulating phase.
The renormalized band gap is introduced as an efficient order parameter for the
transition from a band insulator. We also present the spectral properties of the
ionic Hubbard model, which exhibit characteristic features for both weak and
strong interactions.
\end{abstract}


\maketitle
\section{Introduction}
Strongly correlated electron systems have been one of the most interesting problems
in modern condensed matter physics.
The interest in strongly correlated systems was motivated by experiments
on transition-metal oxides which were inconsistent with the predictions of the conventional band theory.
Since the  argument by Mott 
that the mutual interaction between electrons can cause insulating behavior,
extensive research has been performed on the effects of the interaction in the metal-insulator transition.~\cite{Imada1998}
It is practically impossible to include all the degrees of freedom in strongly correlated systems,
and the standard approach has been to solve theoretical models constructed with essential ingredients.
The Hubbard model (HM) is one of the most popular models in strongly correlated 
systems.~\cite{Hubbard1963} It includes only two essential components:
electron hoppings and local Coulomb interactions.
Although the model seems extremely simplified, it has successfully described the metal-insulator transition
caused by the mutual interaction between electrons.

The ionic Hubbard model (IHM), an extended version of the HM,
was proposed to explain the neutral-ionic transition
in the quasi-one-dimensional charge-transfer organic materials.~\cite{Strebel1970, Hubbard81, Nagaosa86_1, Nagaosa86_2,
Nagaosa86_3}
Unlike the original HM,
this model is an insulator in the absence of the mutual Coulomb interaction.
It enables one to examine an interesting insulator-insulator transition,
from a band insulator (BI) under weak interaction
to a Mott insulator (MI) under strong interaction.
Accordingly the transition nature in the IHM is expected to
differ significantly from the usual metal-insulator transition in the HM.

A more interesting feature of this model
is the possibility of a nontrivial intermediate state,
sandwiched between the two insulating phases.
Extensive studies of the intermediate state in the IHM
have been carried out in various spatial dimensions.
In the infinite dimensions, which can be treated exactly by  
the dynamical mean-field theory (DMFT),~\cite{Georges96}
the metallic phase is observed as an intermediate phase for the weak alternating potential,
whereas a direct insulator-insulator transition is shown under a strong staggered potential.~\cite{Garg2006, Craco2008, Byczuk2009}
The same conclusion has been obtained by a recent study using a coherent potential approximation.~\cite{Hoang2010}
In two dimensions, on the other hand, 
there has been some controversy as to the nature of the intermediate phase
in the IHM.
Quantum Monte Carlo calculations~\cite{Paris2007, Bouadim2007} showed that
the metallic phase exists, as is observed in the infinite dimensions.
In contrast, the cellular dynamical mean-field theory (CDMFT)~\cite{Kancharla2007} 
as well as the variational cluster approach\cite{Chen2010}
predicts a bond-ordered insulating phase as an intermediate state.

In one dimension, which is our main interest in this paper,
the possibility of an insulating intermediate state was suggested by the bosonization 
method,~\cite{Fabrizio99}
which predicts 
a spontaneously dimerized insulating (SDI) phase between BI and MI 
phases.
Many subsequent interesting works for the one-dimensional (1D) IHM have been 
reported.~\cite{Kampf03, Manmana04, Torio2001, Wilkens2001, Otsuka2005, Aligia2005}
Particularly density matrix renormalization group (DMRG) calculations have 
confirmed the existence of an SDI phase for intermediate interaction strengths.~\cite{Kampf03, Manmana04}
However, the critical properties of the transition between BI and SDI phases
obtained by DMRG calculations turned out not to be consistent with those of two-dimensional (2D)
Ising universality class predicted in a bosonization method. 
For the transition between SDI and MI phases, even the critical interaction
strength has not been identified clearly from finite-size scalings.

The main purpose of this paper is to investigate in detail the nature of
transitions in the 1D IHM at half filling.
In order to achieve this 
we use the CDMFT,~\cite{Kotliar2001} 
which is one of the cluster extensions~\cite{Maier05} of DMFT.
While a single site is chosen to construct the self-consistent equation
in the DMFT, the CDMFT picks up a cluster composed of several sites in the 
spatial dimension of the system.
This makes it possible to include short-ranged spatial fluctuations inside the
cluster, which are expected to be important in low-dimensional systems.
The CDMFT turns out to be a very efficient method even in one dimension,~\cite{Bolech03, Capone04, Koch2008, Go2009}
the worst case for a mean-field theory.
In particular, it has been shown that the CDMFT of the 1D 
HM yields 
excellent agreement with the Bethe ansatz exact solution.%
\cite{Capone04, Koch2008, Go2009}
Such a good agreement in turn reinforces our expectation of
an accurate description by the CDMFT method of the 1D IHM
for which no exact solution is available. 
We employ the exact diagonalization method as an impurity solver to
study the ground state of the 1D IHM.
The exact diagonalization method is powerful since it deals with all the quantum
fluctuations on an equal footing, although it limits the number of sites inside the
cluster as well as that of bath sites.
Recent studies on correlated systems show that
single-site or cluster DMFT combined with the exact diagonalization gives reliable results for finite temperatures as well.~\cite{Capone2007, Liebsch2009, Liebsch2011} 

We calculate local quantities such as staggered charge density and double occupancy,
which have been reported to be evaluated accurately in the CDMFT approach.~\cite{Capone04, Koch2008}
We also introduce a renormalized band gap defined by the self-energy corrected band gap and
demonstrate that it plays the role of convenient order parameter for the BI phase.
Any indication of the transition from an intermediate phase to an MI phase is not observed 
in the local quantities,
such as the abrupt change in slope which was reported in the cellular dynamical
mean-field study of the two-dimensional IHM.~\cite{Kancharla2007}
In order to gain further insight into the transition nature we finally 
focus on the spectral properties of the model.
Computed spectral weights reveal the spin-charge separation in an MI phase,
which is characteristic of the 1D system.
Such spin-charge separation has been reported
by earlier studies through cluster extensions of the DMFT 
only for the 1D HM.~\cite{Senechal00, Go2009}
The transition from an intermediate phase to an MI phase exhibits rather a 
crossoverlike behavior, which may be a reason for the difficulty in obtaining
the transition point clearly in earlier numerical investigations. 

This paper is organized as follows.
In Sec.~\ref{sec.MM}, we describe the IHM 
and introduce briefly the procedure of the CDMFT.
The numerical results and discussions are presented in Sec.~\ref{sec.results}.
We give a summary in Sec.~\ref{sec.summary}

\section{\label{sec.MM}Model and Method}

The Hamiltonian of the 1D IHM is given by
\begin{align}
    H=& -t\sum_{\langle i,j \rangle\sigma}
    (c^{\dagger}_{i\sigma} c^{}_{j\sigma}+ \mathrm{H.c.})
    - \mu \sum_{i\sigma} n_{i\sigma}\nonumber\\
    &+ U \sum_i n_{i\uparrow}n_{i\downarrow}
    - \frac{\Delta}{2} \sum_{i\in A} n_{i\sigma} + \frac{\Delta}{2} \sum_{i\in B} n_{i\sigma},
\end{align}
where $c^{\dagger}_{i\sigma}$ ($c^{}_{i\sigma}$) creates
(destroys) an electron with spin $\sigma$ at the $i$th site and 
$n^{}_{i\sigma}{\equiv}c^{\dagger}_{i\sigma} c^{}_{i\sigma}$. 
The hopping of electrons is allowed only between the nearest neighbors.
The parameters $t$, $U$, and $\mu$ are
the hopping amplitude, the
on-site Coulomb repulsion, and the chemical potential, respectively.
Throughout the paper we will represent all the energies in units of $t$. 
The system is composed of 
two alternating sublattices $A$ and $B$. The potential energy
difference between nearest neighbors is $\Delta(>0)$ and every site belonging
to the sublattice $A$ ($B$) has lower (higher) potential energy  
by $\Delta/2$ than the chemical potential. If $\Delta$ vanishes
this model is restored to the original HM. We set
the chemical potential to be half the Coulomb repulsion $(\mu = U/2)$ 
in order to maintain the half filling of electrons over the system.

We use the CDMFT to study the 1D IHM.
The infinite lattice is reduced to a cluster of size $N_c$
which hybridizes with the self-consistent electronic bath sites within the CDMFT.
We employ the exact diagonalization method as an impurity solver.
In order to obtain the cluster self-energy, we map the lattice model to the impurity Hamiltonian,
\begin{align}
    H_{\mathrm{imp}} = & \nonumber
 \sum_{\mu\nu\sigma}E_{\mu\nu} c^{\dagger}_{\mu\sigma}c_{\nu\sigma}
    + U \sum_\mu n_{\mu\uparrow}n_{\mu\downarrow}
\\ & 
    + \sum_{\mu l\sigma} ( V^{}_{\mu l\sigma}  a^{\dagger}_{l\sigma}c^{}_{\mu\sigma} +V^{*}_{\mu l\sigma}  c^{\dagger}_{\mu\sigma}a^{}_{l\sigma} )
    + \sum_{l\sigma} \epsilon^{}_{l\sigma}a^{\dagger}_{l\sigma} a^{}_{l\sigma},
\end{align}
where $\mu,\nu=1,2,\ldots,N_c$ are the cluster site indices and
$l=1,2,\ldots,N_b$ label the bath sites.
The matrix $\hat{E}$ contains 
the hoppings, the staggered potential, and the chemical potential inside the cluster, which is given explicitly by
\begin{align}
\hat{E} \equiv \left(
\begin{matrix}
-\mu - \Delta/2 & -t & 0  & \cdots & 0  \\
-t & -\mu + \Delta/2 & -t & \cdots & 0 \\
0 & -t & -\mu - \Delta/2  & \cdots & 0 \\
\vdots & \vdots & \vdots  & \ddots & \vdots &  \\
0 & 0 & 0 & \cdots & -\mu + \Delta/2
\end{matrix}
\right) .
\end{align}
[The circumflex over a symbol $(\hat{\phantom{\_}})$ represents a $N_c \times N_c$ matrix.]
The parameters $\{\epsilon_{l\sigma}\}$ and $\{V_{\mu l \sigma}\}$, which respectively denote
the bath energy levels and hybridization strengths with the clusters,
are determined from the imposed self-consistency conditions.
From the impurity Hamiltonian we compute the cluster Green function $\hat{G}$ 
as well as the cluster self-energy, $\hat{\Sigma}^c=\hat{\mathcal{G}}^{-1} - \hat{G}^{-1}$,
where $\hat{\mathcal{G}}$ is the Weiss field describing the noninteracting bath.

The Green function of momentum $\tilde{k}$ in reduced Brillouin zone is given by
\begin{align}
    \hat{G} (\tilde{k},i\omega_n)
    = \left[ \hat{M} - \hat{t}(\tilde{k}) - \hat{\Sigma}^c(i\omega_n) \right]^{-1},
    \label{eq.Gk}
\end{align}
where 
$M_{\mu\nu} {\equiv} (i\omega_n-E_{\mu\nu}) \delta_{\mu\nu}$, 
$\hat{t}(\tilde{k})$ is the Fourier transform of the hopping matrix, and
$\omega_n=(2n+1)\pi/\beta$, $n=1,2,\ldots, N_\mathrm{max}$ are fictitious Matsubara frequencies. 
Here we used $\beta=100$ and $N_\mathrm{max}=400$
and
omitted the spin index $\sigma$ for simplicity.
The local lattice Green function 
$\hat{G}_\mathrm{loc}$ is then determined by
\begin{align}
    \hat{G}_\mathrm{loc}(i\omega_n)
    = \sum_{\tilde{k}} \hat{G} (\tilde{k},i\omega_n).
\end{align}
The new Weiss field $\hat{\mathcal{G}}_\mathrm{new}$ is obtained through the self-consistent equation,
\begin{align}
    \hat{\mathcal{G}}_{\mathrm{new}}^{-1}(i\omega_n) = \hat{G}^{-1}_\mathrm{loc}(i\omega_n) + \hat{\Sigma}^c(i\omega_n),
\end{align}
the best fit of which produces new bath parameters $\{\epsilon_{l\sigma},V_{\mu l\sigma} \}$. 
The above procedure is repeated until the convergence is reached.
For a more detailed procedure, we refer the reader to earlier works.\cite{Go2009, Go2010}

It is known that even the cluster of $N_c=2$, which is the minimal size describing the system within the CDMFT approach,
is sufficient to obtain an accurate estimation of the local quantities in 
the 1D HM.~\cite{Capone04}
Further,
we have observed that in the 1D IHM 
 the local quantities obtained by the calculations for various combination of $N_c$ and $N_b$ are not significantly dependent on the choice of $N_c$ and $N_b$,
although larger $N_c$ has some tendency to improve the momentum resolution of spectral weights. 
The number of sites in the impurity Hamiltonian is practically limited 
by $N_c+N_b\lesssim 12$.
We use $N_c=4$ and $N_b=8$ for the presentations of most quantities investigated
in this paper, and other choices of $N_c$ and $N_b$ will be specified.
\section{\label{sec.results}Results and Discussion}
\subsection{\label{subsec.local}Local quantities}

In order to  examine how the system evolves with the variation of the local
interaction we first calculate two local
quantities: staggered charge density and double occupancy.
The staggered charge density is defined by the difference between 
the number densities at two sublattices, $n_A{-}n_B$, where  
the sublattice number densities can be calculated as
\begin{align}
    n_{\alpha}
    &\equiv
    \frac{2}{N_c} \sum_{\mu \in \alpha} \sum_\sigma \langle   n_{\mu\sigma}  \rangle \quad (\alpha = A, B),
\end{align}
with the angular brackets being the expectation value over the ground state of the impurity Hamiltonian.
We also calculate the double occupancy defined by
\begin{align}
D \equiv \frac{1}{N_c} \sum_\mu \langle  n_{\mu\uparrow}n_{\mu\downarrow}
\rangle, 
\end{align}
which is known to be a convenient measure in the transition from metal to 
an MI phase.

\begin{figure}[t]
    \centering \subfigure[]{\includegraphics[angle=-90,width=7cm]{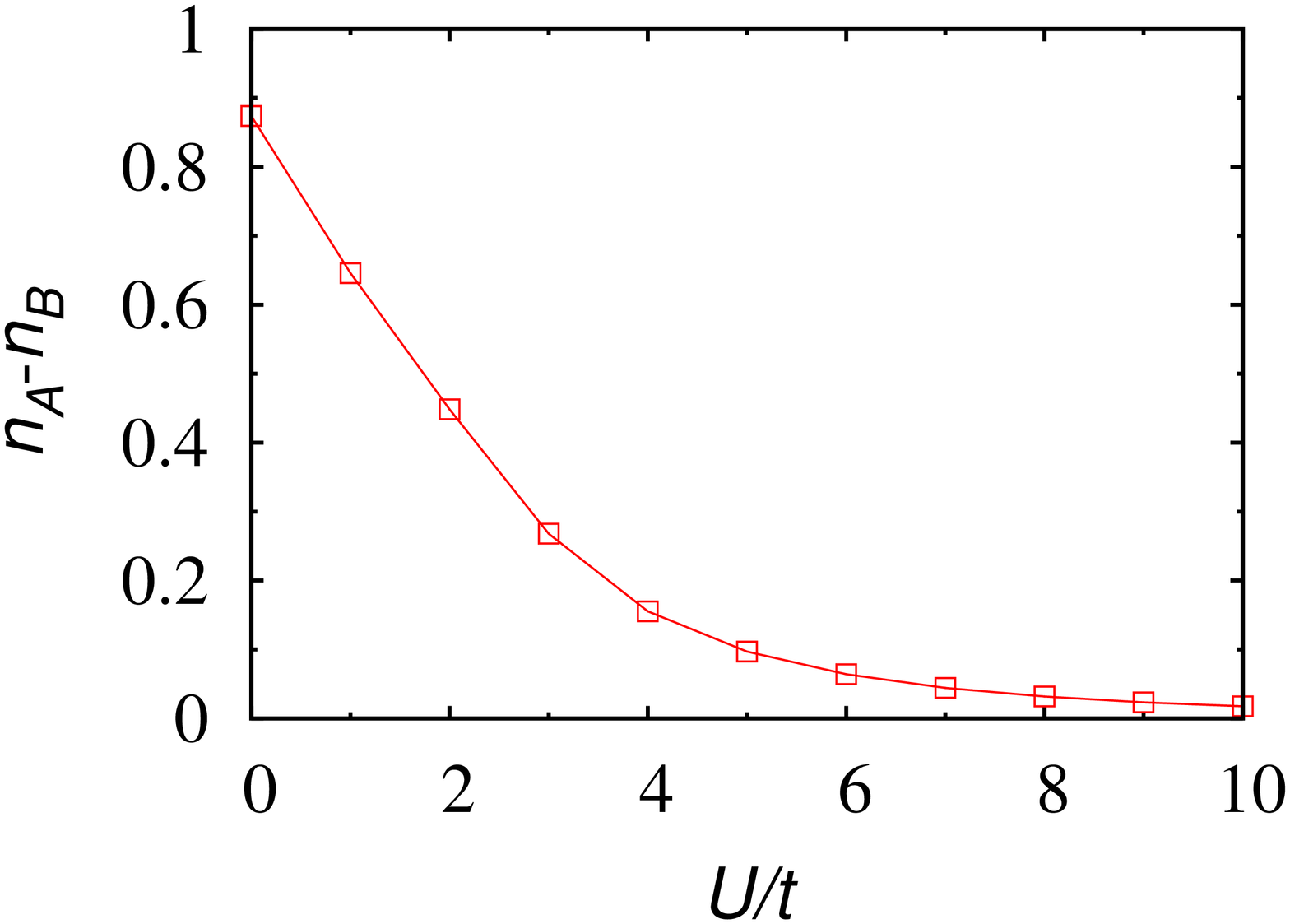}}\\
    \centering \subfigure[]{\includegraphics[angle=-90,width=7cm]{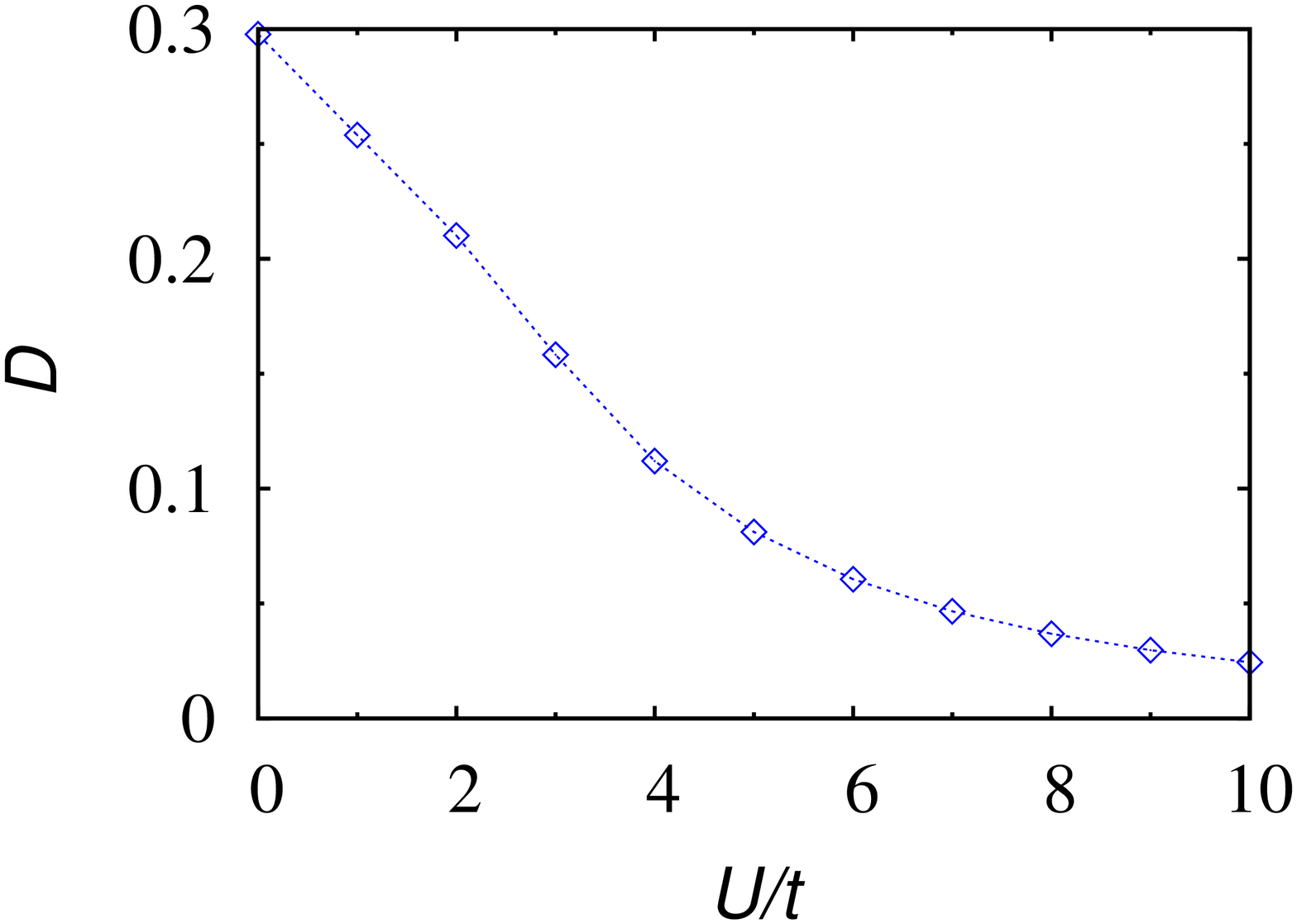}}
    \caption{\label{fig.stag}
        (Color online) (a) Staggered charge density $n_A{-}n_B$. 
        (b) Double occupancy $D$ as a function of $U/t$ for $\Delta/t=1$. 
Both quantities approach zero monotonically as $U$ is increased, signifying 
that the system exhibits a BI phase for weak interaction while 
an MI is recovered in the limit of strong interaction.
    The lines are merely guides to the eyes.
    }
\end{figure}
In the ionic limit ($\Delta\gg t,U$), it is energetically favorable that 
all the electrons are in the sublattice $A$,
producing unity of the staggered charge density.
As $U$ is increased,
the energy cost of two electrons to stay in the same site becomes large, 
reducing the staggered charge.
In the strong-coupling limit ($U\gg t,\Delta$),
the staggered charge is expected to approach zero. 

In the case that the system has  neither the interaction nor 
the staggered potential $(U=\Delta=0)$,
the double occupancy is $1/4$ since
every site has an equal possibility to be occupied independent of spin.
The presence of 
the staggered potential in the system tends to
increase the double occupancy.
In contrast,
the repulsion between electrons makes the doubly occupied sites 
less favorable and results in the reduction of the double occupancy.

The features of two quantities described  in the above reasoning 
are generally consistent with our numerical results in 
Fig.~\ref{fig.stag}, which shows (a) the staggered 
charge density and (b) the double occupancy as a function of $U$.
The staggered charge is relatively large in the weak interaction regime 
and monotonically decreases with increasing $U$; it reveals that the system
exhibits a BI phase for weak interactions and that 
an MI behavior turns up in the regime of strong interactions.
In our numerical work we have also confirmed that
the system with large $\Delta$ shows larger staggered charge density 
over the whole region although the increment is reduced for stronger
interactions (not shown).
The analysis of double occupancy has drawn a similar conclusion.
In the noninteracting system ($U=0$) the double occupancy is larger than 
the free-electron value 1/4 and the system lies in a BI phase.
The double occupancy is reduced monotonically by the increase of $U$,
approaching zero, as is expected in an MI. 
Monotonic decrease in both the staggered charge density and the double occupancy
demonstrates that the increase of the interaction strength drives the system
from a BI into an MI. 
It is of interest to note that no abrupt change is observed in the variation
of both quantities with an increase of the interaction strength.

\begin{figure}[t]
    \centering \includegraphics[angle=-90,width=8cm]{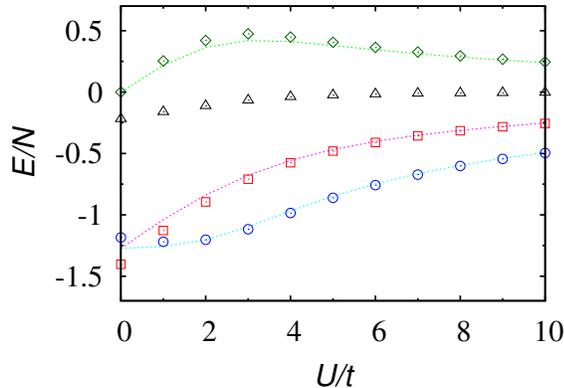}
    \caption{\label{fig.energy}
    (Color online)
    Energy densities $\Delta/t=1$ as a function of $U$.
    The total, kinetic, potential, and correlation energy densities are 
denoted by squares($\square$), circles($\bigcirc$), triangles($\triangle$), and
diamonds($\Diamond$), respectively.  
      The corresponding energy densities in the 1D HM
    are given as lines for comparison except for the potential one which is zero
in the HM.
    }
\end{figure}
\subsection{\label{subsec.energy} Energy densities}
In order to gain more insight on the transition between BI and MI, 
we compute the energy density of the system.
The ground-state energy density of the system is calculated as
\begin{align} \nonumber
\frac{E}{N}
&=
\epsilon_K + \epsilon_\Delta + \epsilon_U, \nonumber\\
\epsilon_K &\equiv
\frac{2}{\beta} \sum_{n} \sum_{\tilde{k}} \left[
   \frac{1}{N_c}{\rm Tr}
\left\{
\hat{t}(\tilde{k})
\hat{G}(\tilde{k},i\omega_n)
\right\}
\right],\nonumber\\
\epsilon_\Delta &\equiv \Delta (n_A - n_B),\nonumber\\
\epsilon_U &\equiv U D,
    \label{eq.cenergy}
\end{align}
where $\epsilon_K$, $\epsilon_\Delta$, and $\epsilon_U$ denote kinetic, potential, and correlation
energy densities, respectively.
The factor 2 in $\epsilon_K$ comes from spin degeneracy.
Computed energy densities illustrate which contribution plays a dominant 
role in each phase.
In Fig.~\ref{fig.energy} we have plotted various contributions to the 
energy density of the 1D IHM
along with those of the 1D HM for comparison.
Since the HM does not have staggered potential, the contribution from
potential energy is zero and is not plotted in the figure.

In the region of weak interactions ($U\lesssim 2t$), we can see that both the
kinetic and the correlation energy densities are higher than those of the
standard HM. The energy gain in the potential contribution
compensates for the increase in other energy densities, demonstrating the
BI nature of the system in this region.
The increase of the interaction strength lessens such a tendency gradually.
In the strong-interaction region ($U\gtrsim4t$)
all the energy-density contributions are almost the same as those of the HM
and the energy contribution from the staggered potential is almost zero,
signifying that the system is in an MI phase.
As in the local quantities examined in the previous subsections, all the energy
densities display continuous variations with the increase of interaction
strength.

\subsection{\label{subsec.localDOS} Local density of states}
\begin{figure}[t]
\includegraphics[angle=-90,width=7cm]{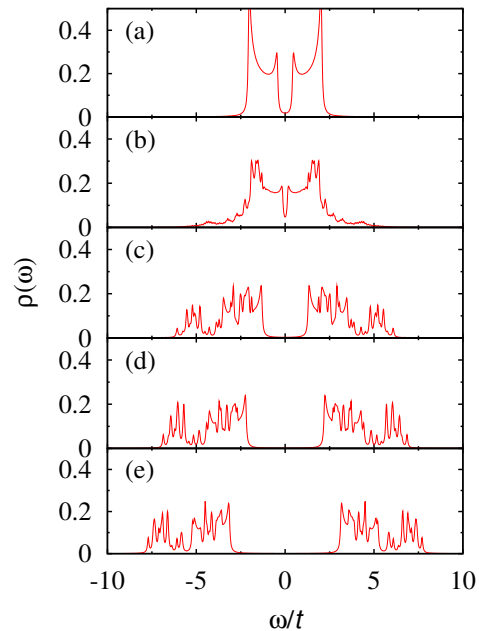}
    \caption{\label{fig.dos}
        (Color online)
    Local density of states $\rho(\omega)$ for $\Delta/t{=}1.0$
        and $U/t{=}0.3, 2.4, 6.0, 8.0,$ and $10.0$ from top to bottom.
 In the weak-coupling regime the system displays a two-band structure 
        and the gap between the two bands is reduced gradually with the increase
of $U$.
In the strong-coupling limit, on the other hand,
the two bands move away from the Fermi level and each band seems to be split into 
two subbands.
        A broadening factor $\varepsilon{=}0.05$ is used.
    }
\end{figure}

The local density of states (LDOS) provides more detailed information on 
the single-particle properties.
Within the CDMFT approach, the LDOS is given by
\begin{align}
    \rho(\omega) = \sum_k A(k, \omega),
\end{align}
where $A(k, \omega)$ is a spectral weight with the energy $\omega$ and 
the momentum vector $k$ in the full Brillouin zone.
We restore the translational symmetry broken in the CDMFT formalism
by the periodization of the Green function for each sublattice 
\begin{align} \label{eq.perG}
    G_{\alpha}(k,\omega)=\frac{2}{N_c}\sum_{\mu,\nu\in \alpha} e^{-ik(\mu-\nu)}
[\hat{G}(k,\omega)]_{\mu\nu} \quad (\alpha = A, B),
\end{align}
where $\hat{G}(k,\omega)$ is given in Eq.~(\ref{eq.Gk}).
Then we can compute the total spectral weight $A(k, \omega)$ as
\begin{align}
    A(k,\omega)=-\frac{1}{\pi} \sum_{\alpha=A,B}
\mathrm{Im}G_{\alpha}(k,\omega+i\varepsilon), 
    \label{eq.sw}
\end{align}
where $G_\alpha$ is the periodized Green function of the sublattice $\alpha$
and $\varepsilon$ is a small broadening factor.
In this work we used $\varepsilon{=}0.05$. 

The LDOS is shown in Fig.~\ref{fig.dos} for several values of $U$.
In the noninteracting system $(U=0)$, the LDOS can be computed analytically 
and is composed of two bands which are separated by a band gap $\Delta$ due to
the staggered potential. 
The CDMFT results generally reproduce the analytical LDOS for a 
noninteracting system as demonstrated in Ref.~\onlinecite{Go08}.
Turning on the interaction 
does not change the LDOS very much from the noninteracting LDOS, and the two
band structure is retained.
For weak interactions the increase of interaction strength reduces
monotonically the band gap around a Fermi level.
Around a certain value $U_0 {\approx} 2.4t$ the LDOS displays a minimum spectral
gap and prominent long tails show up at the outer edges of the bands.
Further increase of $U$ above $U_0$
in turn enlarges the gap between the two bands. 
Each band is apparently split into two subbands and the LDOS is composed of 
four bands for strong interactions.
Such a four-band structure is also observed in the original HM,
where it is caused by
spectral weights concentrated on the spinon-holon continuum.~\cite{Go2009}
In the IHM
the gap between the subbands is distinguished more clearly than in the 
HM and it is found to be proportional to $\Delta$.
We will give more detailed discussions on this topic in the section 
that deals with the spectral weights.

\subsection{\label{sec.gap} Spectral gap and renormalized band gap}
\begin{figure}[t]
    \centering \includegraphics[angle=-90,width=7cm]{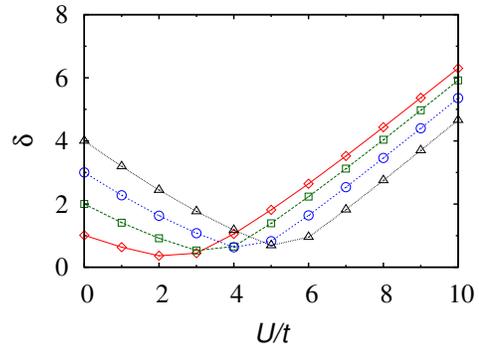}
    \caption{\label{fig.gap}
    (Color online)
        Spectral gaps $\delta$ as a function of the interaction strength $U$ for various $\Delta/t$. 
    The spectral gaps for the system with $\Delta/t=1, 2, 3$, and $4$
        are denoted by diamonds($\Diamond$), squares($\square$), circles($\bigcirc$), and triangles($\triangle$), respectively.
With the increase of $U$ the spectral gap decreases for weak interactions while
it grows larger in the regime of strong interactions.
    }
\end{figure}
For a quantitative analysis of the gap around a Fermi level,
we investigate the spectral gap $\delta$, which is defined as 
the energy difference between the highest filled and the lowest 
empty levels in the LDOS. 
Since the IHM always displays a minimum gap at 
$k{=}\pi/2$ 
the spectral gap can be conveniently obtained from the spectral weight 
at the Fermi point $k{=}\pi/2$ within the CDMFT.
Since the system has a particle-hole symmetry, 
we estimate the spectral gap $\delta$ to be twice the difference 
between the Fermi level and the peak of $A(k,\omega)$ closest to it for 
positive $\omega$. 
The measured gap is plotted as a function of the interaction
strength $U$ for various $\Delta$ in Fig.~\ref{fig.gap}. 
The overall behaviors of the spectral gap are consistent with those observed in
the LDOS.
The spectral gap is equal to $\Delta$ in the noninteracting system ($U{=}0$). 
It is reduced from the noninteracting value $\Delta$ by weak interactions,
reaches a minimum at $U=U_0$, and increases with $U$ in the regime of strong
interactions above $U_0$.
Our CDMFT results of the system with various $\Delta$ in Fig.~\ref{fig.gap} 
have shown that $U_0$ increases with the bare band gap $\Delta$.

In a BI quasiparticle excitations are well defined and we can
interpret the spectral gap as the energy difference between 
the lowest quasielectron and quasihole excitations.
By expanding the self-energy for each sublattice around the Fermi level
we can estimate the position of the poles of the Green function
on the real frequency axis near the Fermi level.
We then define the {\em renormalized band gap} $\Delta_\mathrm{ren}$
by the difference between the poles of the Green's function of the
sublattice $A$ and $B$ which are the closest to the Fermi level, which is
expected to be the same as the spectral gap in a BI phase.
By definition
$\Delta_\mathrm{ren}$ equals a bare band gap $\Delta$
in the noninteracting system
since the self-energy vanishes
and the poles of the Green function are identical to the bare dispersion of 
the system.
Indeed $\Delta_\mathrm{ren}$ is a band gap with a self-energy correction.
On the other hand, in correlated phases such as an MI, 
the quasiparticle excitations are not well defined and the discrepancy 
between $\delta$ and $\Delta_\mathrm{ren}$ will show up.
We can thus expect that the comparison of the renormalized band gap
$\Delta_\mathrm{ren}$ 
with the spectral gap $\delta$ will be a good parameter for determining whether the 
system is in a BI phase or not.

\begin{figure}[t]
    \centering \includegraphics[angle=-90,width=7cm]{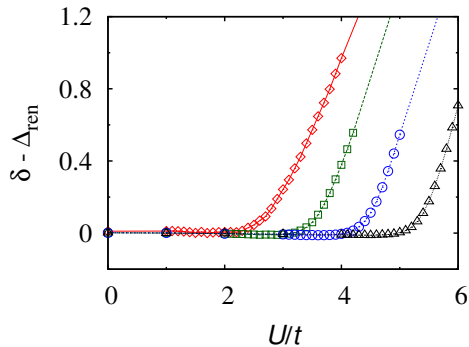}
    \caption{\label{fig.ren}
        (Color online) Difference between the spectral gap $\delta$ and 
the renormalized band gap $\Delta_\mathrm{ren}$ for various values of 
$\Delta/t$.
        The same symbols are used as in Fig.~\ref{fig.gap}.
    }
\end{figure}

Within the CDMFT we have computed and plotted the difference between the 
spectral gap $\delta$ and $\Delta_\mathrm{ren}$ in Fig.~\ref{fig.ren}.
We can see that it is zero for $U{=}0$,
which is guaranteed from the definition of the renormalized band gap.
It is remarkable that the difference remains zero 
over a finite region of $U$ below a certain critical value $U_\mathrm{c1}$. 
Above $U_\mathrm{c1}$ the discrepancy between
$\delta$ and $\Delta_\mathrm{ren}$ turns out to grow rapidly,
signifying that an MI phase or another correlated phase 
emerges for $U>U_\mathrm{c1}$.
For $\Delta=t$, the CDMFT yields $U_{c1}\approx2.2t$;
this is comparable to, although slightly smaller than, the critical values 
obtained in the existing works,
$U_\mathrm{c1} \sim 2.67t$ from the DMRG study~\cite{Manmana04} and
$U_\mathrm{c1} \sim 2.3t$ from the effective model in the strong-coupling limit.~\cite{Tincani2009}
We compare $U_\mathrm{c1}$ and  $U_0$ for various $\Delta$ in Table~\ref{tab.uco}.
It is found that two interaction strengths correlate very much with each other
and $U_0$ is always slightly higher than $U_\mathrm{c1}$, which is also consistent 
with existing works.~\cite{Manmana04,Kampf03} 

\begin{table}[t]
    \caption{\label{tab.uco}
        Comparison of the critical interaction strength $U_\mathrm{c1}$ below which 
the system is in a BI phase and
 $U_0$ at which the system shows a minimum spectral gap.
All the energies are given in unit of $t$.
    }
    \begin{ruledtabular}
    \begin{tabular}{c c c}
        $\Delta$    &   $U_0$    &   $U_\mathrm{c1}$\\
        \hline
        $1.0$  &   $2.4$  &   $2.2$  \\
        $2.0$  &   $3.4$  &   $3.2$  \\
        $3.0$  &   $4.4$  &   $4.1$  \\
        $4.0$  &   $5.3$  &   $5.0$  \\
    \end{tabular}
    \end{ruledtabular}
\end{table}

\subsection{\label{sec.SW} Spectral weights}
\begin{figure}[t]
    \centering \subfigure[]{\includegraphics[angle=-90,width=6cm]{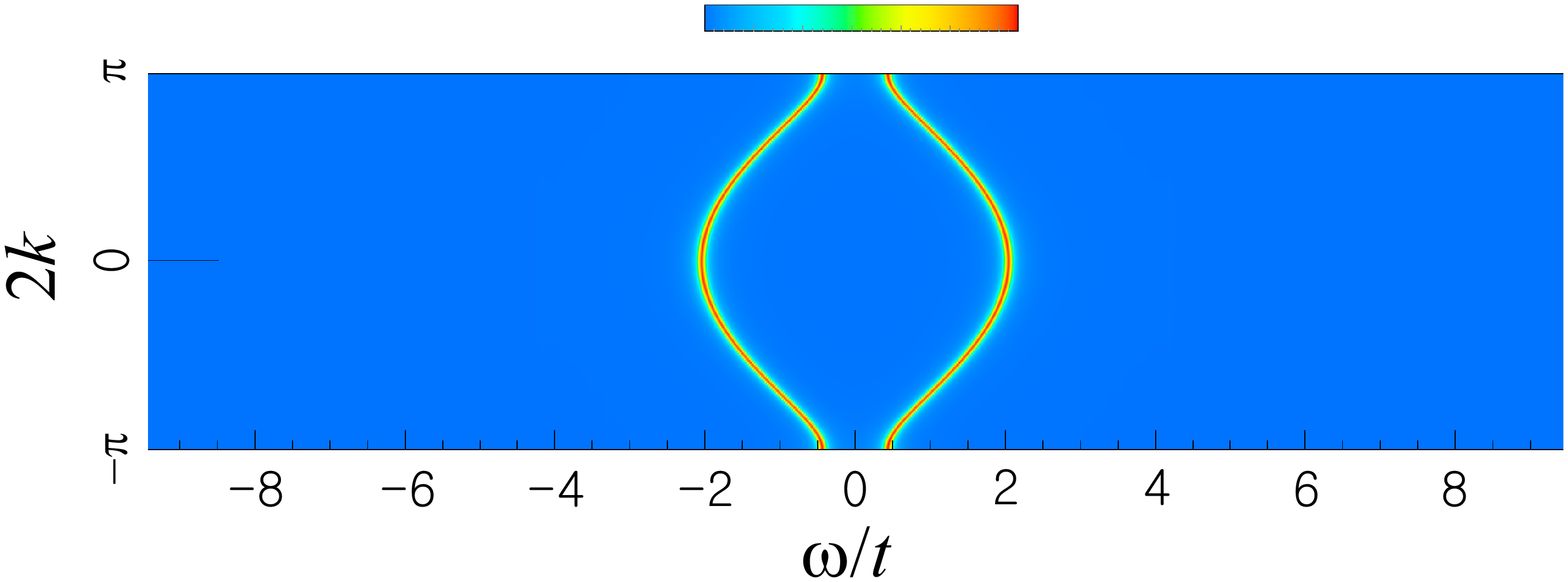}}\\
    \centering \subfigure[]{\includegraphics[angle=-90,width=6cm]{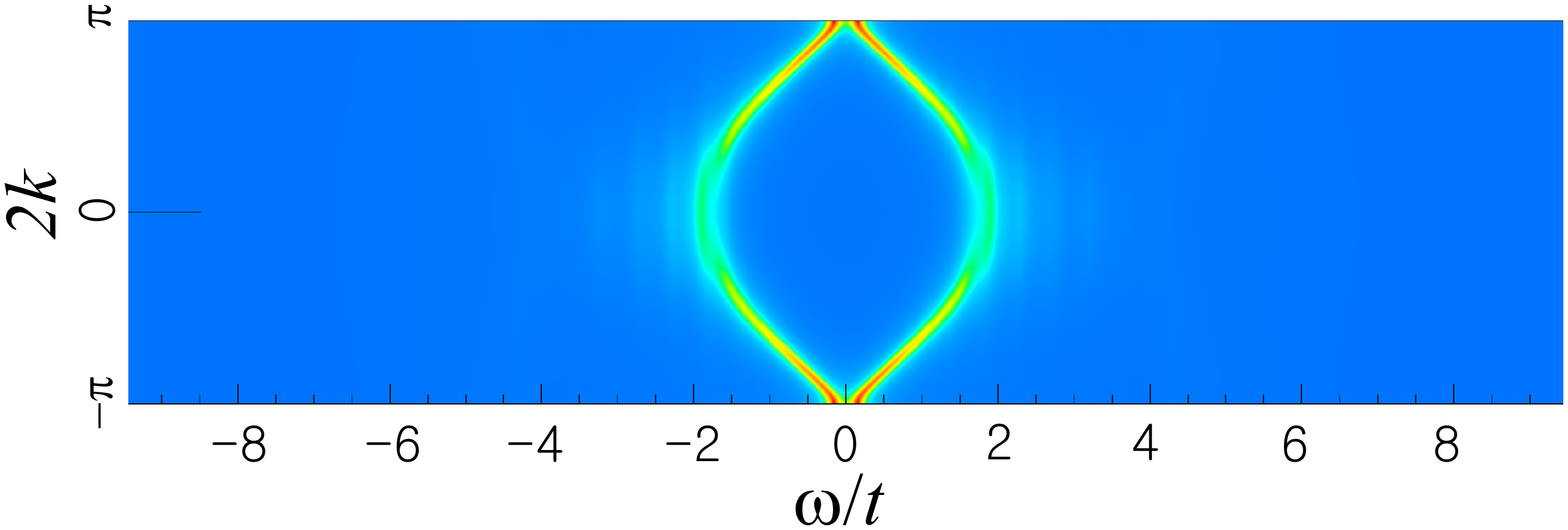}}\\
    \centering \subfigure[]{\includegraphics[angle=-90,width=6cm]{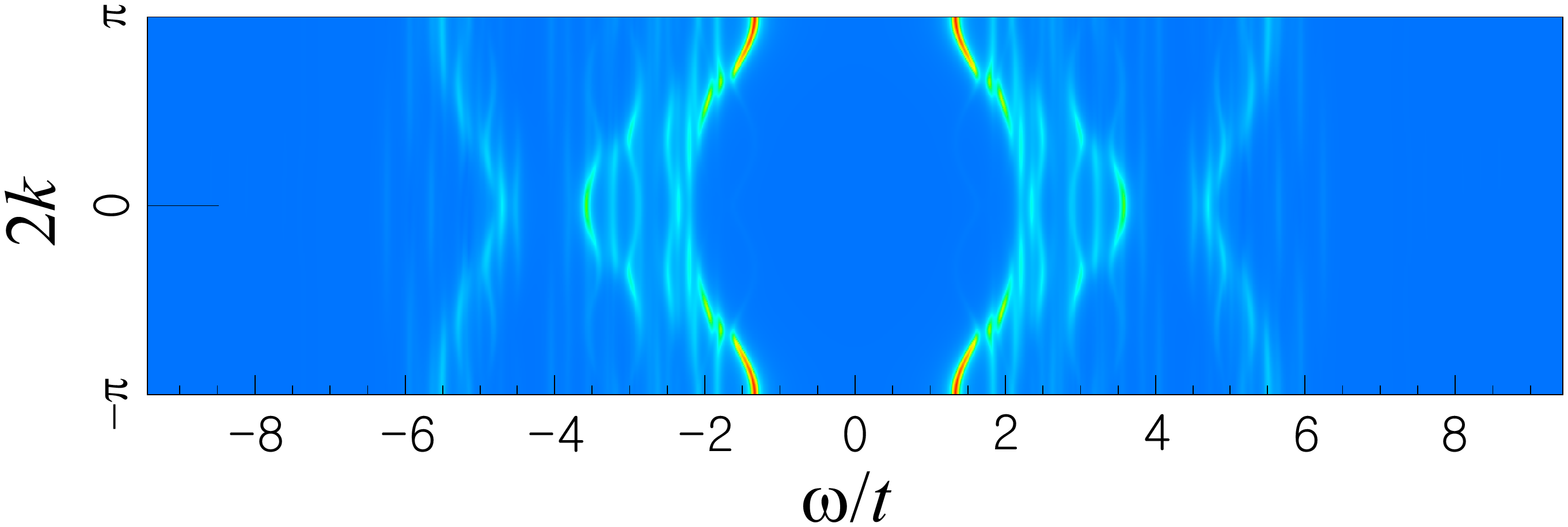}}\\
    \centering \subfigure[]{\includegraphics[angle=-90,width=6cm]{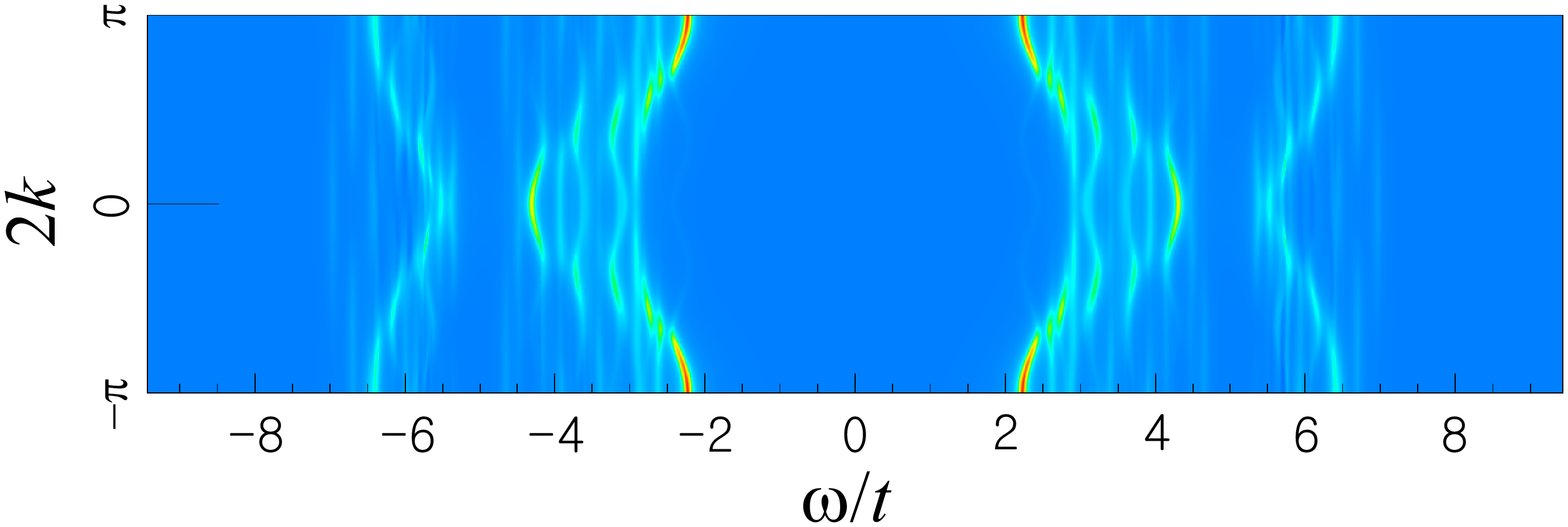}}
    \caption{\label{fig.sw}
    (Color online)
        Spectral weights for $\Delta/t=1.0$ and (a) $U/t=0.3$; (b) $U/t=2.2$; (c) $U/t=6$; and (d) $U/t=8$.
    We used a cluster of size $N_c{=}6$ with $N_b{=}6$ bath sites. 
        Since the unit cell of the IHM is twice that of the 
original HM, the first Brillouin zone is reduced to $-\pi/2<k<\pi/2$.
    For better resolution the spectral weights are rescaled according to their
maximum values in each plot.
}
\end{figure}

A useful quantity for demonstrating
momentum-resolved information on the correlation effects in the system
is a total spectral weight $A(k,\omega)$ given in Eq.~(\ref{eq.sw}). 
This is computed from periodized sublattice Green functions 
in Eq.~(\ref{eq.perG}) and we plot $A(k,\omega)$ for various interaction
strength $U$ in Fig.~\ref{fig.sw}.

For weak interaction below $U_\mathrm{c1}$ two quasiparticle bands are sharply 
defined in $A(k,\omega)$ as illustrated in Fig.~\ref{fig.sw}(a) for $U{=}0.3t$. 
Such quasiparticle bands which are separated by a band gap result in 
two-band structure of the LDOS which is similar to that of
a noninteracting system. 
In this regime only the gap between the two bands and the band
widths are renormalized by weak interactions.

In Fig.~\ref{fig.sw}(b) we can observe that around $U_\mathrm{c1}$
the single-particle dispersion begins to be broadened particularly 
around the zone center $k{=}0$, which is a signature of the transition 
to a correlated phase from a BI phase. 
The broadening around the zone center is a source for the appearance of long
tails at the outer edges of the bands in the LDOS.
Here it is of interest to note that the broadening shape of the dispersion
closely resembles the spin-charge separation observed in 
the 1D HM.
The spinon (holon) is a fractionalized
excitation which carries only spin (charge) but no charge (spin).
In the 1D HM, the Bethe ansatz solution gives the dispersion of
the exotic excitations. The assembled dispersions construct the
spectral weight and we can observe some prominent dispersions such as
the spinon and the holon branches. 
In the 1D HM most weights of $A(k, \omega)$ are
concentrated on the spinon branch under relatively weak
interaction and more weights transfer to the holon branches and other
accessible regions of higher energies as the 
interaction strength is increased.~\cite{Go2009} 
Similar weight transfer is observed in the spectral weights of 1D IHM as shown in Fig.~\ref{fig.sw}.

\begin{figure}[t]
    \centering \subfigure[]{\includegraphics[angle=-90,width=6cm]{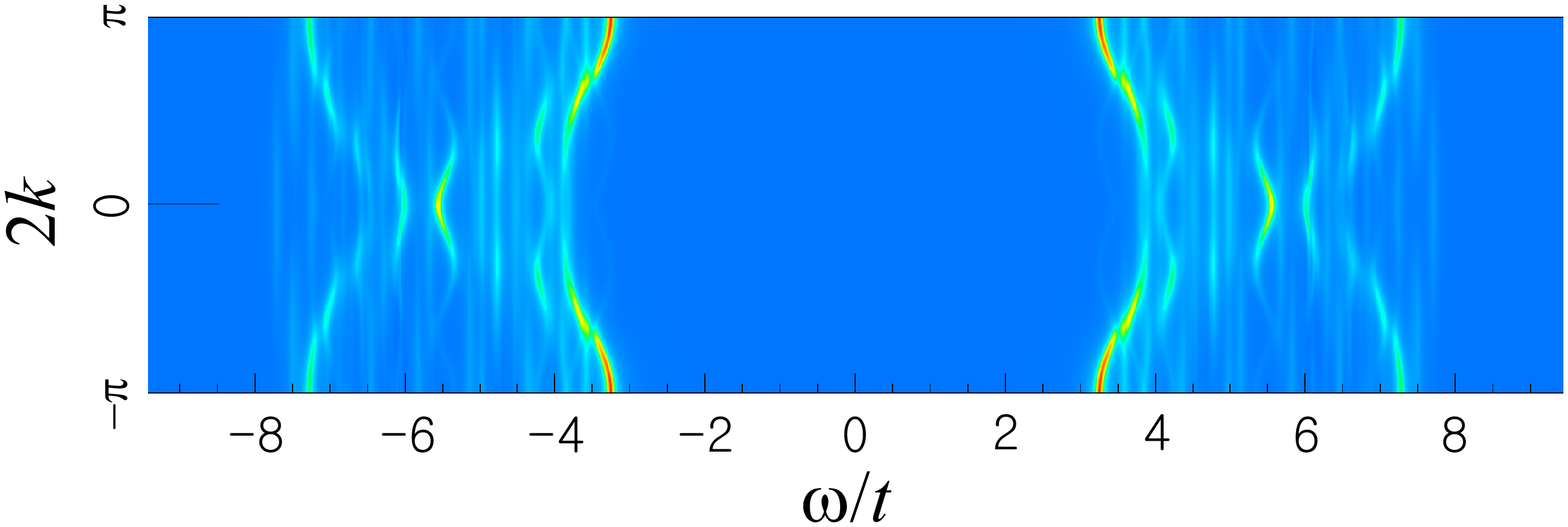}}\\
    \centering \subfigure[]{\includegraphics[angle=-90,width=6cm]{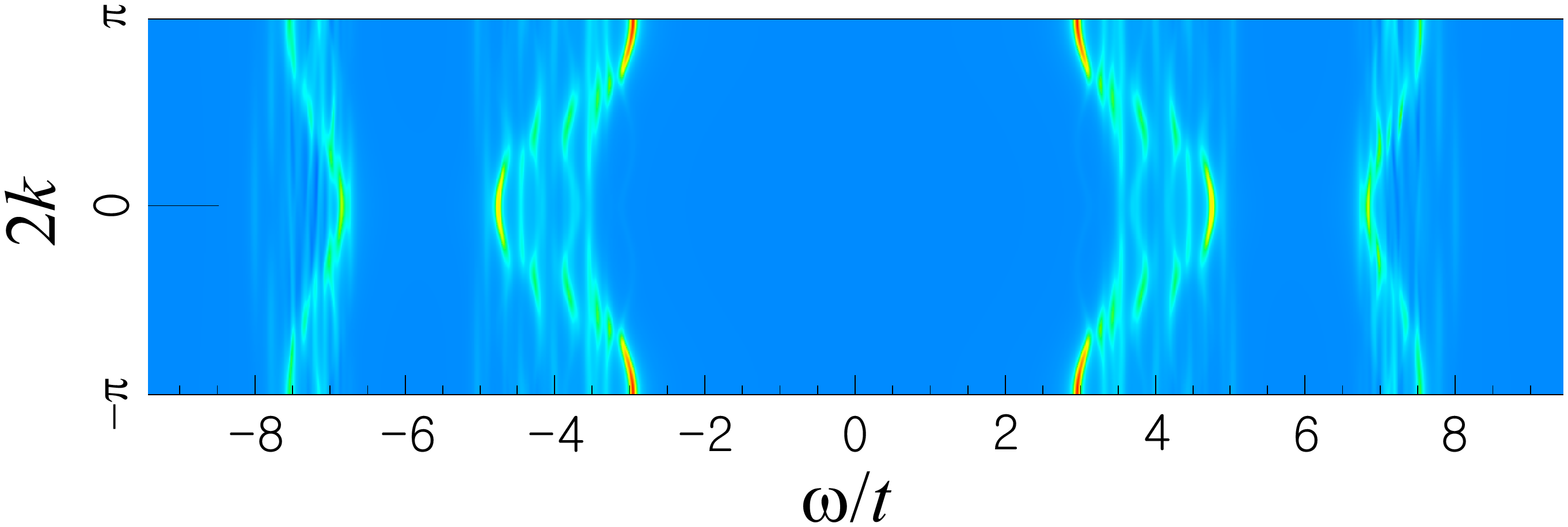}}\\
    \centering \subfigure[]{\includegraphics[angle=-90,width=6cm]{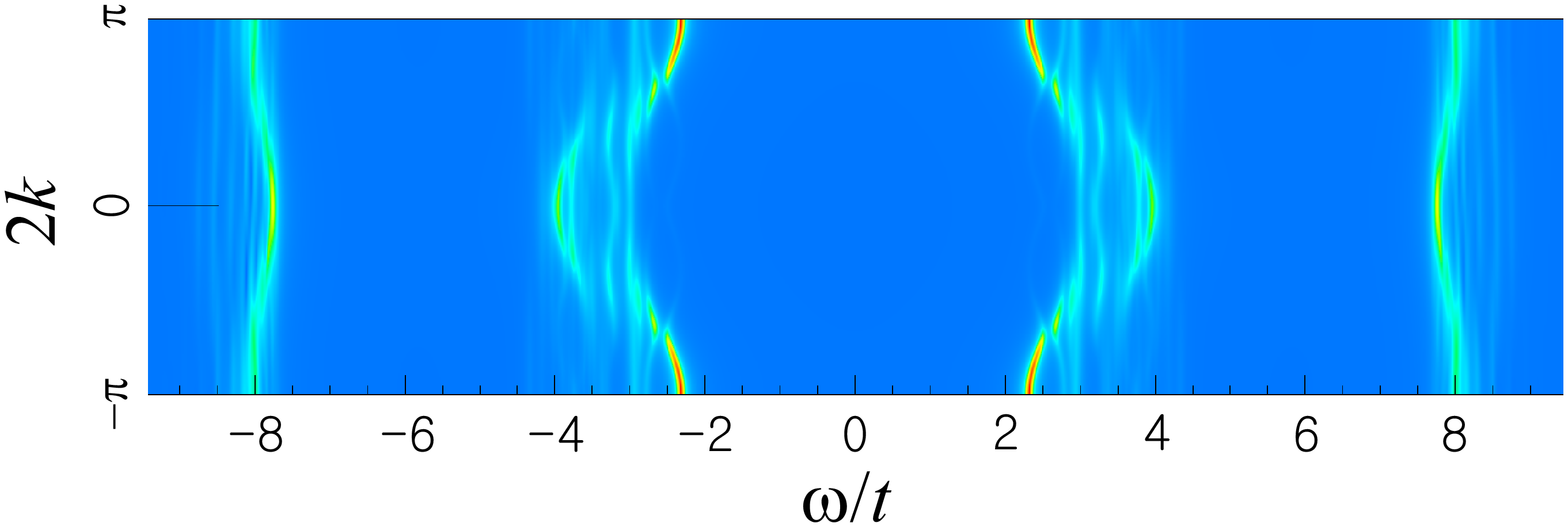}}
    \caption{\label{fig.sw2}
    (Color online)
        Spectral weight for $N_c=6$, $N_b=6$, $U/t=10.0$, (a) $\Delta/t=0$, (b) $\Delta/t=2$, and (c) $\Delta/t=4$.
    }
\end{figure}

Although the spectral weights of the two models are quite similar to 
each other in the strong interaction regime,
a remarkably different feature also arises from the presence of 
staggered potential.
In the absence of the staggered potential two holon branches for $k{>}0$ and
for $k{<}0$ cross each other at the zone center slightly above $\omega{=}U/2$, 
and extends as secondary holon branches in the other region,
as is reproduced in Fig.~\ref{fig.sw2}(a).~\cite{Go2009}
The spectral weights for $\Delta/t=2$ and $4$, which are shown in
Figs.~\ref{fig.sw2}(b) and \ref{fig.sw2}(c), demonstrate marked gaps at the crossing points
of the holon branches.
The fact that the gap width is proportional to $\Delta$ 
also supports that the degeneracy of the holon branches at $k{=}0$ is lifted by
the staggered potential.
Accordingly, the dispersion displays four well-separated bands, 
yielding the characteristic
four-band structure observed in the LDOS for large $U$ in Fig.~\ref{fig.dos}. 
We also note that the shift of the spinon and the first holon bands toward the
Fermi level produces some reduction in the Mott gap in the presence of a
staggered potential.

\begin{figure}[t]
    \centering \subfigure[]{\includegraphics[angle=-90,width=7cm]{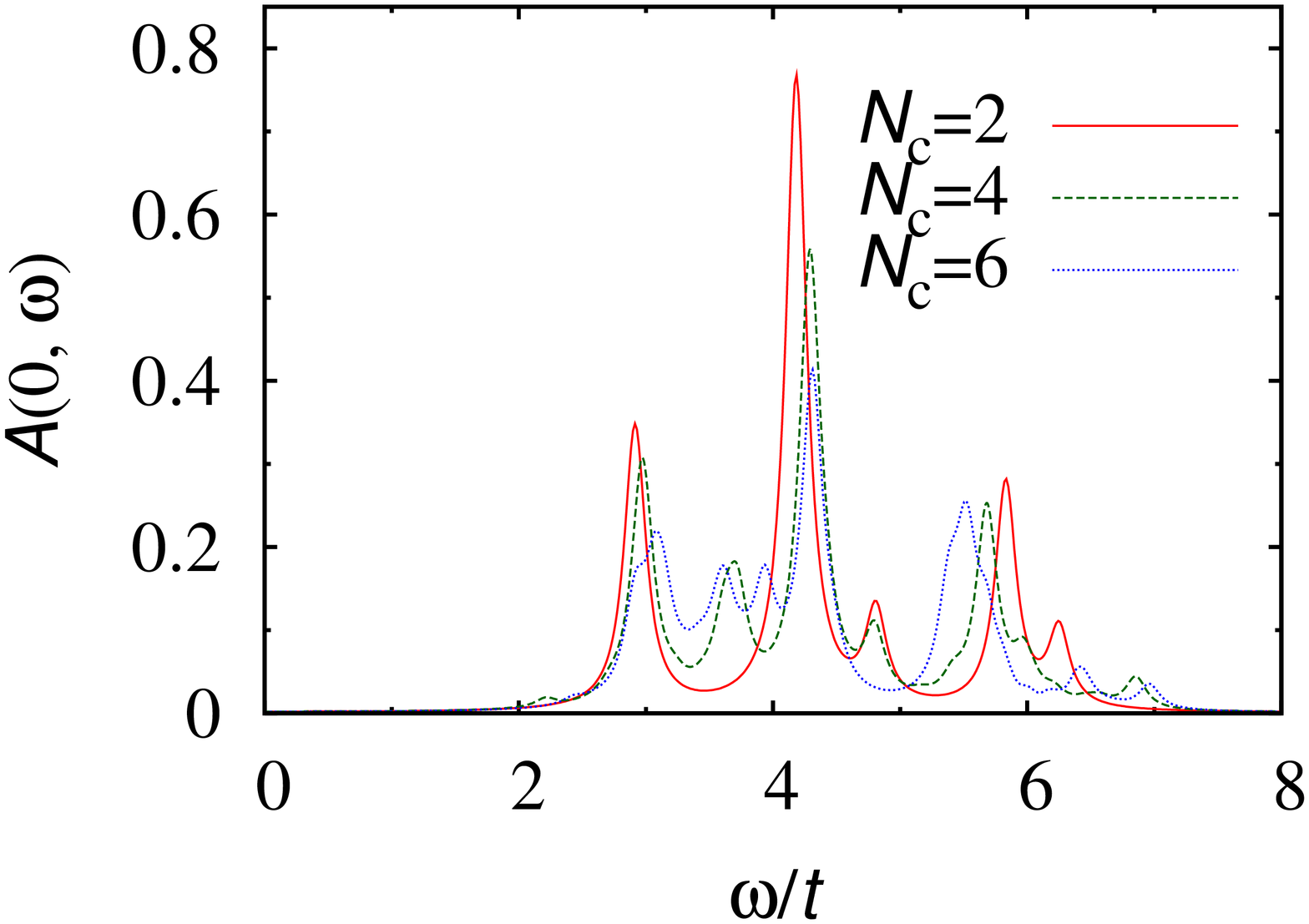}}\\
    \centering \subfigure[]{\includegraphics[angle=-90,width=7cm]{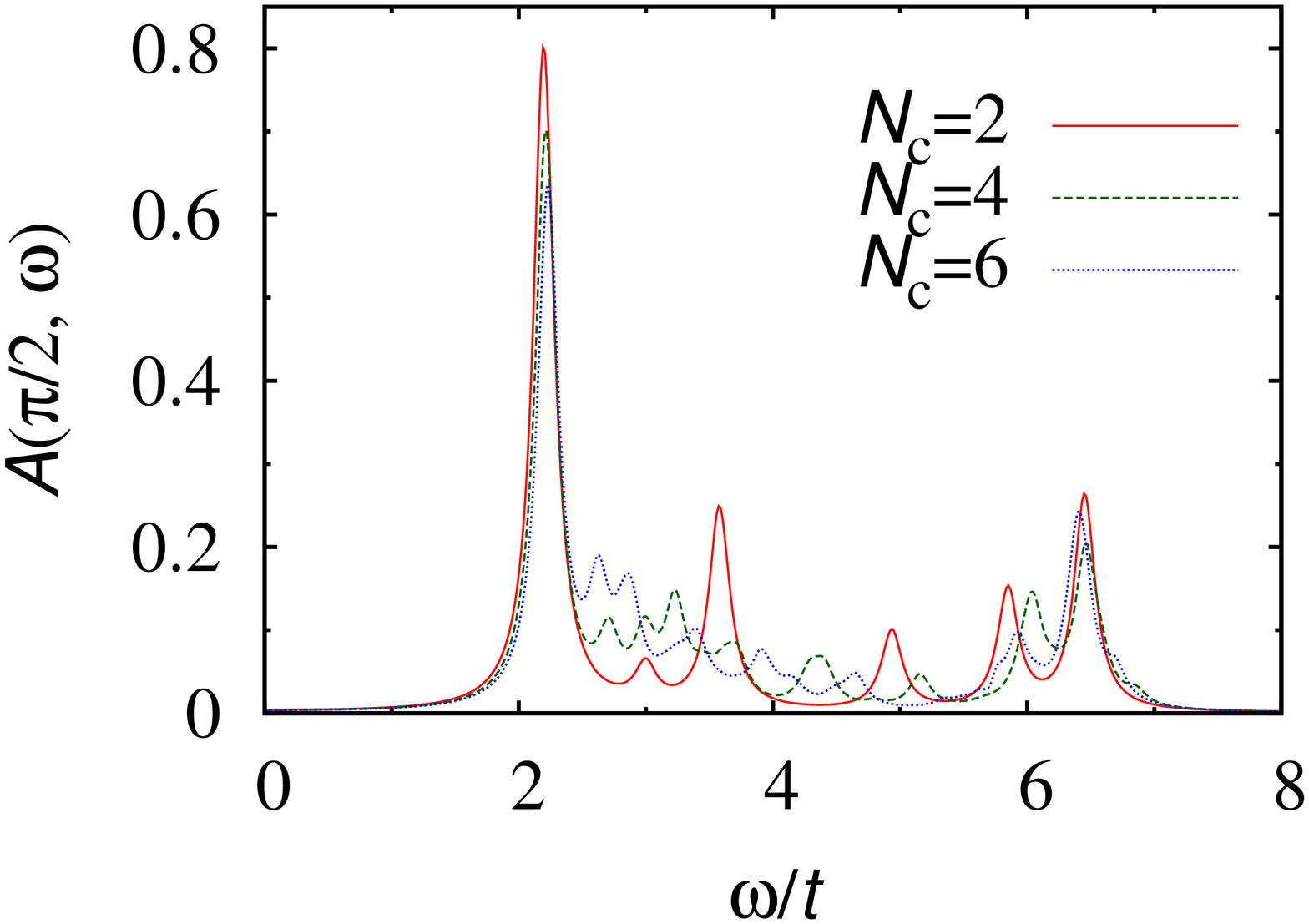}}\\
    \caption{\label{fig.SC}
    (Color online)
        Spectral weight for $\Delta/t=1$, $U/t=8$, (a) $k=0$, and (b) $k=\pi/2$ with different cluster sizes.
    }
\end{figure}

The comparison of the spectral weights with different $N_c$ gives us a good
guide to the understanding of the overall distribution of spectral weights. 
In Fig.~\ref{fig.SC}
we plot the spectral weights on the two momentum points $k{=}0$ and 
$k{=}\pi/2$ with different cluster sizes $N_c{=}2, 4,$ and $6$.
In the plot of spectral weights at the zone center in Fig.~\ref{fig.SC}(a)
we can recognize three prominent peaks around $\omega\approx 3t$, $4.3t$, and
$5.5t$, which are identified as the spinon and the two split holon branches,
respectively.
A similar tendency is demonstrated clearly in the plot of $A(\pi/2,\omega)$ in
Fig.~\ref{fig.SC}(b).
The first peak around $\omega\approx 2t$ corresponds to a merging band of the 
spinon and the lower holon band, while the higher holon band generates 
the second peak around $\omega\approx 6.5t$.
The peaks which seem to be prominent for $N_c{=}2$ tend to be smeared out to a
continuum with the increase of $N_c$.

Since the pioneering bosonization approach~\cite{Fabrizio99} it has been 
generally believed that the 1D IHM shows two successive transitions with an
increase of the interaction strength.
In the CDMFT we have identified the first transition from the BI by examining
the renormalized band gap. However, we are not able to position the second
transition point from the intermediate insulating phase to the MI
in the investigation of the systems with various values of $\Delta$. 
We have not observed any nonanalytic behaviors in local quantities such as 
the staggered charges or the double occupancy, 
in contrast to the 2D IHM where some kinks in the local quantities were 
proposed as a signature of the second phase transition.~\cite{Kancharla2007}
No abrupt change in the spectral properties occurs as the interaction strength
is increased.
Particularly the spectral weights exhibit a rather gradual transition to the
MI. Although the origin of such difficulty in positioning the second phase
transition is not clear, we believe that it is related to rather slow
falloff of the bond-order parameter around the second transition in the DMRG
study.~\cite{Manmana04}

\section{\label{sec.summary}Summary}
We have investigated the one-dimensional half-filled ionic Hubbard
model at zero temperature using the cellular dynamical mean-field
theory. 
We have computed the staggered charge density
and the double occupancy. 
Both quantities display monotonic decrease with an increase of the interaction
strength, signifying that the system evolves from a band insulator to a Mott
insulator.
The energy-density analysis shows that the potential energy gain occurs for weak
interactions while the system with strong interactions gives almost the same
energy contributions as in the Hubbard model.
The phase boundary of a band insulating phase has been determined by the 
comparison of the renormalized band gap with the spectral gap.
Around the phase boundary we have also observed a minimum of the spectral gap
for various strengths of staggered potential.
We have calculated the
spectral weights and analyzed the detailed structure of spin-charge
separation by analogy with the one-dimensional Hubbard model. 
The staggered potential turns out to produce a gap at the crossing point of two
holon branches, which is proportional to a band gap parameter.

\section*{ACKNOWLEDGMENTS}
This research was supported by Basic Science Research Program through the National Research Foundation of Korea (Grant No. 2010-0010937) funded by the Ministry of Education, Science and Technology.

\bibliography{IHM}
\end{document}